\documentclass[pra,twocolumn,superscriptaddress]{revtex4-2}
\usepackage{float,graphicx,multirow, amsmath,color,dsfont}
\usepackage{amsmath,bm}
\usepackage{amssymb,mathrsfs,dsfont}
\usepackage{amssymb,mathbbol}
\usepackage{amsfonts}
\usepackage{mathrsfs}
\usepackage{float}
\usepackage{xcolor,hyperref}
\definecolor{darkblue}{rgb}{0,0.0.1,0.3}
\definecolor{darkred}{rgb}{0.6,0.1,0}
\hypersetup{colorlinks,breaklinks,
	linkcolor=darkred,urlcolor=darkblue,
	anchorcolor=darkred,citecolor=
	darkred,pdfauthor=ChSh, pdftitle=ChMoSh.Noisy}
\usepackage{tikz}
\usepackage{mathbbol}
\usepackage{float}
\usepackage[normalem]{ulem}%For bibliography 
\newcommand{\ie}{\textit{i}.\textit{e}.}

\begin{document}

	\title{ 
		Continuous variable quantum teleportation in a dissipative environment:
		Comparison of non-Gaussian operations before and after noisy
		channel}
	
	\author{Chandan Kumar}
	\email{chandan.quantum@gmail.com}
	\affiliation{Department of Physical Sciences,
		Indian
		Institute of Science Education and
		Research Mohali, Sector 81 SAS Nagar,
		Punjab 140306 India.}
	\author{Mohak Sharma}
	\email{mohak.quantum@gmail.com}
	\affiliation{Department of Physical Sciences,
		Indian
		Institute of Science Education and
		Research Mohali, Sector 81 SAS Nagar,
		Punjab 140306 India.}
	\author{Shikhar Arora}
	\email{shikhar.quantum@gmail.com}
	\affiliation{Department of Physical Sciences,
		Indian
		Institute of Science Education and
		Research Mohali, Sector 81 SAS Nagar,
		Punjab 140306 India.}
	\begin{abstract}

		We explore the relative advantages in continuous-variable quantum teleportation when non-Gaussian operations, namely, photon subtraction, addition, and catalysis, are performed before and after interaction with a noisy channel. We generate the resource state for teleporting unknown coherent and squeezed vacuum states using two distinct strategies: (i) Implementation of non-Gaussian operations on TMSV state before interaction with a noisy channel, (ii) Implementation of non-Gaussian operations after interaction of  TMSV state with a noisy channel. The results show that either of the two strategies could be more beneficial than the other depending on the type of the non-Gaussian operation, the initial squeezing of the TMSV state, and the parameters characterizing the noisy channel. This strategy can be utilized to effectively improve the efficiency of various non-Gaussian continuous variable quantum information processing tasks in a dissipative environment.

	\end{abstract}
	\maketitle
	%%%%%%%%%%%%%%%%%%%%%%%%%%%%%%%%%%
	\section{Introduction}

	Quantum teleportation~\cite{Bennett,Vaidman}  plays a central role in manipulating quantum states, long-distance quantum key distribution~\cite{long-distance,repeater}, and distributed quantum computing~\cite{dqc}. In continuous variable (CV) quantum teleportation, two-mode squeezed vacuum (TMSV) state is most commonly   employed as a resource state~\cite{bk-1998}.  With the current experimental techniques, it is not possible to obtain a highly squeezed state~\cite{15dB}, which puts an upper bound  (less than one) on the teleportation fidelity.
	To overcome this limitation, one can resort to non-Gaussian (NG) operations, which have been shown to enhance the performance of not only quantum teleportation~\cite{tel2000,Akira-pra-2006,Anno-2007,tel2009,wang2015,catalysis15,catalysis17,tele-arxiv} but also various other protocols, such as quantum key distribution~\cite{qkd-pra-2013,qkd-pra-2018,qkd-pra-2019,qk2019, zubairy-pra-2020} and  quantum metrology~\cite{gerryc-pra-2012,josab-2012,braun-pra-2014,josab-2016,pra-catalysis-2021,crs-ngtmsv-met,metro-thermal-arxiv,ngsvs-arxiv}.

	Dissipation is a detrimental factor in the performance of various QIP tasks, which can originate due to interaction with a noisy channel. It affects quantum protocols based on Gaussian~\cite{pablo-prl-2008,sandeep-pra-2010,Paulina-scripta-2015,rcga} as well as non-Gaussian states~\cite{nonGdiss,nonGdiss-1,nonGdiss-2}.  Since interaction with a noisy channel is  inevitable in a real world scenario, it is important to find strategies for mitigating or lessening its detrimental effects. 
	
	In this article, we systematically analyze whether we should perform NG operations before or after the interaction with a noisy channel in order to reduce the ill effect of dissipation on the fidelity of quantum teleportation. We consider   a realistic scheme for implementing three distinct  NG operations, namely, photon subtraction (PS), photon addition (PA), and photon catalysis (PC)  on a two-mode state [Fig.~\ref{ng}]. In the first strategy, we first perform  NG operations on the TMSV state followed by interaction with the noisy channel [Fig.~\ref{noisycheck}(a)]. In the second strategy, we first let the TMSV state interact with the noisy channel and then perform the NG operations [Fig.~\ref{noisycheck}(b)]. The resource states generated using these two different strategies are utilized for the quantum teleportation of unknown input coherent  and squeezed vacuum states. 
	
	We compare the  fidelity for teleporting  an input coherent state using resource states generated via   two different strategies. For PS operation at high temperatures, we find that it is advantageous to perform PS operation before the interaction with the noisy channel for the initial period of interaction. 
	However, after a certain period of interaction with the noisy channel, this trend reverses, \ie, it is advantageous to perform PS operation after the interaction with the noisy channel. The same behavior is also noted for the PC operation.
	At low  temperature, the difference in the fidelity arising due to the implementation of the PS operation before or after the interaction with the noisy channel is negligible irrespective of the squeezing of the original TMSV state. However, whether we should perform PC operation before or after the interaction with the noisy channel for performance improvement depends on the squeezing and the time of interaction.  Similar trends are observed for the teleportation of the squeezed state.

	Our study  extends and complements earlier studies carried out in the context of teleportation of input coherent state~\cite{betterresource,catalysis17}. For instance, in Ref.~\cite{betterresource}, it has been examined whether it is beneficial to perform ideal single photon subtraction on both the modes of TMSV state before or after the noisy channel. Similarly, in Ref.~\cite{catalysis17}, it has been investigated whether it is advantageous to perform single photon catalysis on both the modes of TMSV state before or after lossy channel.

	The layout of this paper is as follows. In Sec.~\ref{Preliminaries}, we provide a brief introduction to   CV systems and its phase space formalism. 
	We then move on to Sec.~\ref{twostrategies}, where we briefly discuss   noisy channel and the considered NG operations. We also describe the two different strategies considered for resource state preparation.  In Sec.~\ref{comparison}, we numerically compare the fidelity using the two distinct resource states through a series of plots. In Sec.~\ref{sec:conc}, we summarise the main points of the article and its relevance in future studies.

	%%%%%%%%%%%%%%%%
	\section{ Preliminaries of CV systems}\label{Preliminaries}
	We consider an $n$-mode radiation field described by a column vector of quadrature operators~\cite{arvind1995,Braunstein,adesso-2007,weedbrook-rmp-2012,adesso-2014}
	\begin{equation}\label{eq:columreal}
		\hat{ \xi}  =(\hat{ \xi}_i)= (\hat{q_{1}},\,
		\hat{p_{1}}, \dots, \hat{q_{n}}, 
		\, \hat{p_{n}})^{T}, \quad i = 1,2, \dots ,2n.
	\end{equation}
	The canonical commutation relation  between the quadrature operators can be expressed in an elegant form as   
	\begin{equation}\label{eq:ccr}
		[\hat{\xi}_i, \hat{\xi}_j] = i \Omega_{ij}, \quad (i,j=1,2,...,2n),
	\end{equation}
	where we have taken $\hbar=1$ and $\Omega$ is the symplectic form matrix of order 2$n$ $\times$ 2$n$   given by
	\begin{equation}
		\Omega = \bigoplus_{k=1}^{n}\omega =  \begin{pmatrix}
			\omega & & \\
			& \ddots& \\
			& & \omega
		\end{pmatrix}, \quad \omega = \begin{pmatrix}
			0& 1\\
			-1&0 
		\end{pmatrix}.
	\end{equation}
	An $n$-mode CV system can also be expressed by $n$ pairs of 
	photon annihilation and creation 
	operators $\hat{a}_i\, \text{and}\, 
	{\hat{a}_i}^{\dagger}$ ($i=1,2,\, \dots\, ,n$), which are related to the quadrature operators through the relation
	\begin{equation}\label{realtocom}
		\hat{a}_i=   \frac{1}{\sqrt{2}}(\hat{q}_i+i\hat{p}_i),
		\quad  \hat{a}^{\dagger}_i= \frac{1}{\sqrt{2}}(\hat{q}_i-i\hat{p}_i).
	\end{equation}

	In this article, we shall be using the beam splitter operation, $B_{ij}(T)$, and the two mode squeezing operation, $S_{ij}(r)$, which come under the class of symplectic transformations. Below, we describe the transformation of the quadrature operators $\hat{\xi} = (\hat{q}_{i}, \,\hat{p}_{i},\, \hat{q}_{j},\,
	\hat{p}_{j})^{T}$ by these operations:
	\begin{equation}\label{beamsplitter}
		\hat{\xi}' = B_{ij}(T) \hat{\xi} = \begin{pmatrix}
			\sqrt{T} \,\mathbb{1}_2& \sqrt{1-T} \,\mathbb{1}_2 \\
			-\sqrt{1-T} \,\mathbb{1}_2& \sqrt{T} \,\mathbb{1}_2
		\end{pmatrix} \hat{\xi} ,
	\end{equation}
	\begin{equation}\label{eq:tms}
		\hat{\xi}' = S_{ij}(r) \hat{\xi} = \begin{pmatrix}
			\cosh r \,\mathbb{1}_2& \sinh r \,\mathbb{Z} \\
			\sinh r \,\mathbb{Z}& \cosh r \,\mathbb{1}_2
		\end{pmatrix} \hat{\xi}.
	\end{equation}
	In the beam splitter operation $B_{ij}(T)$, $T$ is the transmissivity of the beam splitter while $r$ is the squeezing parameter in the two-mode squeezing operation $S_{ij}(r)$.
	Further,    $\mathbb{1}_2$ is the identity matrix of size 2, while $\mathbb{Z}$ is the Pauli matrix given by $\text{diag}(1,\, -1)$.

	\subsection{Phase space description}
	
	In the phase space formalism of quantum mechanics, the state of a quantum system $\rho$ is described by a phase space distribution. In this study, we employ the Wigner characteristic function, which is the two dimensional Fourier transform of the Wigner  distribution function. In calculations associated with quantum teleportation, Wigner characteristic function approach leads to mathematical simplicity compared to the usage of the Wigner distribution function. For an $n$ mode quantum system with density operator $\rho$, the Wigner characteristic function is given by
	\begin{equation}\label{wigdef}
		\chi(\Lambda) = \text{Tr}[\hat{\rho} \, \exp(-i \Lambda^T \Omega \hat{\xi})],
	\end{equation}
	where $\Lambda = (\Lambda_1, \Lambda_2, \dots \Lambda_n)^T$ with  
	$\Lambda_i = (\tau_i, \sigma_i)^T \in \mathcal{R}^2$.  
	For CV systems, there are states whose Wigner distribution function takes on a Gaussian form. These states are termed Gaussian states and can be completely specified by the first and second-order moments of the quadrature operators. The first moment of an $n$-mode CV system is defined by the displacement vector $\bm{d} = \langle  \hat{\xi } \rangle = \text{Tr}[\hat{\rho} \hat{\xi}]$. The second moments are often represented by a $2n\times2n$ matrix termed covariance matrix: 
	\begin{equation}\label{eq:cov}
		V = (V_{ij})=\frac{1}{2}\langle \{\Delta \hat{\xi}_i,\Delta
		\hat{\xi}_j\} \rangle.
	\end{equation}
	Here $\Delta \hat{\xi}_i = \hat{\xi}_i-\langle \hat{\xi}_i
	\rangle$ and $\{\,, \, \}$ denotes the anti-commutator. The expression of Wigner characteristic function(\ref{wigdef}) for a Gaussian state takes the following form~\cite{weedbrook-rmp-2012, olivares-2012}:
	\begin{equation}\label{wigc}
		\chi(\Lambda) =\exp[-\frac{1}{2}\Lambda^T (\Omega V \Omega^T) \Lambda- i (\Omega \bm{d} )^T\Lambda].
	\end{equation}
	To generate the TMSV state, we consider two modes initialized to vacuum state and apply two-mode squeezing operator~(\ref{eq:tms}). Its Wigner characteristic function turns out to be 
	\begin{equation}
		\begin{aligned}
			\chi(\Lambda) =    \exp\big[&- 
			(\tau_1^2+\sigma_1^2+\tau_2^2+\sigma_2^2)\cosh (2r)/4   \\
			&+    (\tau_1 \tau_2-\sigma_1 \sigma_2) \sinh (2r)/2	\big].
		\end{aligned}
	\end{equation}

	\section{Detailed description of the strategies}\label{twostrategies}
	Before describing the two distinct strategies that we intend to investigate, we briefly discuss     noisy channel and  the realistic scheme for implementing NG operations, including PS, PA and PC operations on two-mode system.
	
	\subsection{	Noisy channel}
	We consider that our two-mode system interacts with a  noisy channel or   thermal bath. 
	The time evolution of the density operator $\rho_{12}$ of a two-mode system, when each of the modes interacts locally with independent thermal baths, is given by
	\begin{equation}\label{master_eqn_local}
		\begin{aligned}
			\frac{\partial}{\partial t} \rho_{12} = & \bigg\{ \sum_{i=1,2}
			\frac{\gamma_{i}}{2}(N_{i}+1)(2 \hat{a}_i \rho
			\hat{a}_i^{\dagger}-\hat{a}_i^{\dagger}\hat{a}_i\rho -\rho
			\hat{a}_i^\dagger\hat{a}_i)\\
			& \quad + \frac{\gamma_{i}}{2}N_{i}(2 \hat{a}_i^\dagger \rho
			\hat{a}_i-\hat{a}_i\hat{a}_i^{\dagger}\rho -\rho
			\hat{a}_i\hat{a}_i^\dagger)\bigg\}.
		\end{aligned}
	\end{equation}
	Here $\gamma_{i}$'s and $N_{i}$'s are the decay  constants and the mean photon number of the individual thermal baths, respectively. We will assume that the two baths are identical, \ie, $\gamma_{1}=\gamma_{2}=\gamma$ and $N_{1}=N_{2}=n_{\text{\text{th}}}$. The evolution of the two-mode system given by the master equation~(\ref{master_eqn_local})  can be modeled by a simple setup where each mode impinges on a  beam splitter with a thermal state of mean photon number $n_{\text{\text{th}}}$ in the ancilla mode [see Fig.~\ref{noisycheck}]. The transmissivity $\eta$ of the beam splitter is related to the decay constant $\gamma$ through  the relation $\eta = e^{-\gamma t}$.

	\subsection{Photon subtraction, addition, and catalysis on a two-mode system}
	
	\begin{figure}[h]
		\centering
		\includegraphics[scale=1]{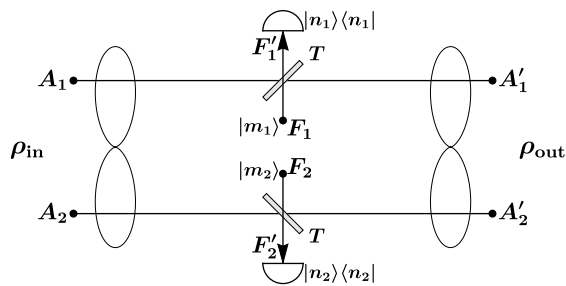}
		\caption{Scheme for photon subtraction, addition, and catalysis on a two mode system.
			Fock states $|m_1\rangle$ and $|m_2\rangle$ are combined with the   modes $A_1$ and $A_2$ of the initial system using beam splitters of transmissivity $T$.
			The output modes corresponding to the ancilla are subjected to conditional measurements of $n_1$ and $n_2$ photons. }
		\label{ng}
	\end{figure}
	
	In Fig.~\ref{ng}, we show the schematic of an experimental setup that can be used to implement NG operations on a two-mode system. These modes are labeled by $A_1$ and $A_2$, and the quadrature operators corresponding to these modes are given by $(\hat{q}_1,\hat{p}_1)^T$ and  $(\hat{q}_2,\hat{p}_2)^T$, respectively. We represent the density operator of the input state by $\rho_{\text{in}}$.    In this study, we   consider  symmetric operations, \ie, NG operations are implemented on both modes. This is because the asymmetric cases (NG operations on only one mode) do not improve the fidelity as compared to the original state~\cite{wang2015,tele-arxiv}. To realize these symmetric NG operations, we mix ancilla mode in Fock state $|m_1\rangle$ ($|m_2\rangle$) with mode $A_1$ ($A_2$) using a beam-splitter of transmissivity $T$. The output ancilla mode $A'_1$ ($A'_2$) is subjected to a conditional measurement of $n_1$ ($n_2$) photons. The simultaneous detection of $n_1$ and $n_2$ photons indicates the successful implementation of NG operations on both modes. 
	We represent the post-measurement state of the two-mode system by the density operator $\rho_{\text{out}}$. As shown in Table~\ref{table1}, different values of the parameters $m_1$, $m_2$, $n_1$, $n_2$, characterize different NG operations \ie, PS, PA and PC operations on the two-mode system. 
	\begin{table}[h!]
		\centering
		\caption{\label{table1}
			Conditions on the values of the parameters $m_1$, $m_2$, $n_1$, $n_2$ for different NG operations.}
		\renewcommand{\arraystretch}{1.5}
		\begin{tabular}{ |c |c |}
			\hline
			Operation & Condition \\
			\hline \hline
			$n$-PS&  $m_1=m_2=0$ and $n_1=n_2=n$ \\ \hline
			$m$-PA&  $m_1=m_2=m$ and $n_1=n_2=0$ \\ \hline
			$n$-PC &  $m_1=m_2=n_1=n_2=n$ \\ \hline
			\hline 
		\end{tabular}
	\end{table}

	\subsection{Two distinct strategies}	  
	
	\begin{figure}[h]
		\centering
		\includegraphics[scale=1]{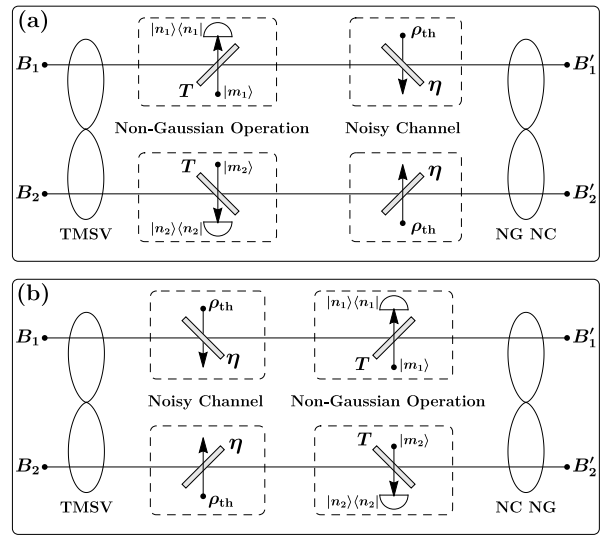}
		\caption{   (a) In the first strategy, NG operations are performed on both modes of the  TMSV state before interaction with a noisy channel (NG~NC). (b) In the second strategy, NG operations are performed only after    the  TMSV state has interacted with a noisy channel (NC~NG).   }
		\label{noisycheck}
	\end{figure}
	
	This investigation aims to reduce the effect of dissipation on the performance of quantum teleportation. To this end, we apply the NG operations   and the noisy channel (NC)   in  two distinct ways  to generate the resource state for quantum teleportation  [see Fig.~\ref{noisycheck}]: 
	\begin{itemize}
		\item[1.]  Performing NG operations  on both modes of the  TMSV state before interaction with a noisy channel (NG~NC).  We denote such operations  for the specific cases of PS, PA, and PC operations by the abbreviations PS~NC, PA~NC and PC~NC, respectively.
		\item[2.]   Performing NG operations on both modes after the  TMSV state has interacted with a noisy channel (NC~NG). We denote such operations for the specific cases of PS, PA, and PC operations by the abbreviations NC~PS, NC~PA, and NC~PC, respectively. 
	\end{itemize}
	The preferred resource state would be the one that yields a higher teleportation fidelity.

	\section{Comparison of the two strategies in quantum teleportation}\label{comparison}
	
	We consider the teleportation of unknown input coherent and squeezed vacuum states via
	Braunstein-Kimble (BK) protocol~\cite{bk-1998}.
	The success probability of the quantum teleportation protocol can be measured
	via fidelity  $F =\text{Tr} [\rho_{\text{in}}\rho_{\text{out}}]$, where $\rho_{\text{in}}$ and $\rho_{\text{out}}$ denote  the density operator of the unknown input state and  the output (teleported) state.  The fidelity can be evaluated in the Wigner characteristic function formalism as follows~\cite{Welsch-pra-2001}:
	\begin{equation}\label{fidex1}
		F =\frac{1}{2 \pi} \int d^2 \Lambda_2  \chi_{\text{in}}(\Lambda_2)
		\chi_{\text{out}}(-\Lambda_2).
	\end{equation}
	The calculation of  fidelity can be simplified by expressing the    Wigner characteristic function of the output state   as~\cite{Marian-pra-2006}:
	\begin{equation}\label{teleported}
		\chi_{\text{out}}(\tau_2,\sigma_2) = \chi_{\text{in}}(\tau_2,\sigma_2) \chi_{A_1' A_2'}(\tau_2,-\sigma_2,\tau_2,\sigma_2),
	\end{equation}
	where  $\chi_{\text{in}}(\tau,\sigma)$ and $\chi_{A_1' A_2'}(\tau_1,\sigma_1,\tau_2,\sigma_2)$ are the  Wigner characteristic functions of the input state and the entangled resource 
	state, respectively.
	The fidelity of teleporting an input coherent state cannot exceed $1/2$ when only classical resources are utilized~\cite{Braunstein-jmo-2000,Braunstein-pra-2001}. Therefore, the value of fidelity going beyond $1/2$ signals the success of quantum teleportation.

	\subsection{ Teleportation of single mode coherent state}
	
	We first compare the  fidelity of teleporting an input coherent state using the   resource states prepared in two different ways as mentioned in Sec.~\ref{twostrategies}. To this end, we first analyze the fidelity as a function of transmissivity $\eta = e^{-\gamma t}$, representing the   time of interaction of the resource state with the thermal bath. The transmissivity $\eta = 1$ corresponds to zero interaction time, while $\eta = 0$ corresponds to  interaction  for an infinite time with the thermal bath. 
	
	\begin{figure*}[t] 
		\centering
		\includegraphics[scale=1]{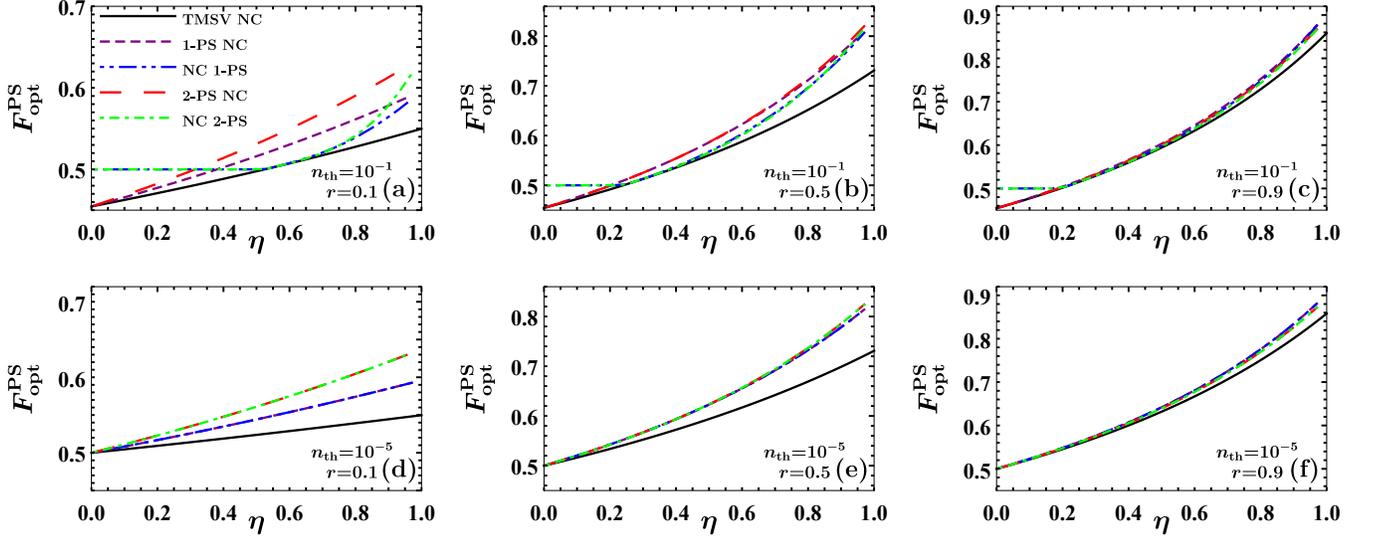}
		\caption{ Optimal fidelity as a function of transmissivity $\eta = e^{-\gamma t}$ for different squeezing values of the TMSV state and   thermal photon number ($n_{\text{th}}$). The   fidelity has been   maximized   with respect to the transmissivity $(T)$ of the beam splitter that implements the photon subtraction operation. }
		\label{ss_grid_optF_vs_eta}
	\end{figure*}
	
	\begin{figure*}[t] 
		\centering
		\includegraphics[scale=1]{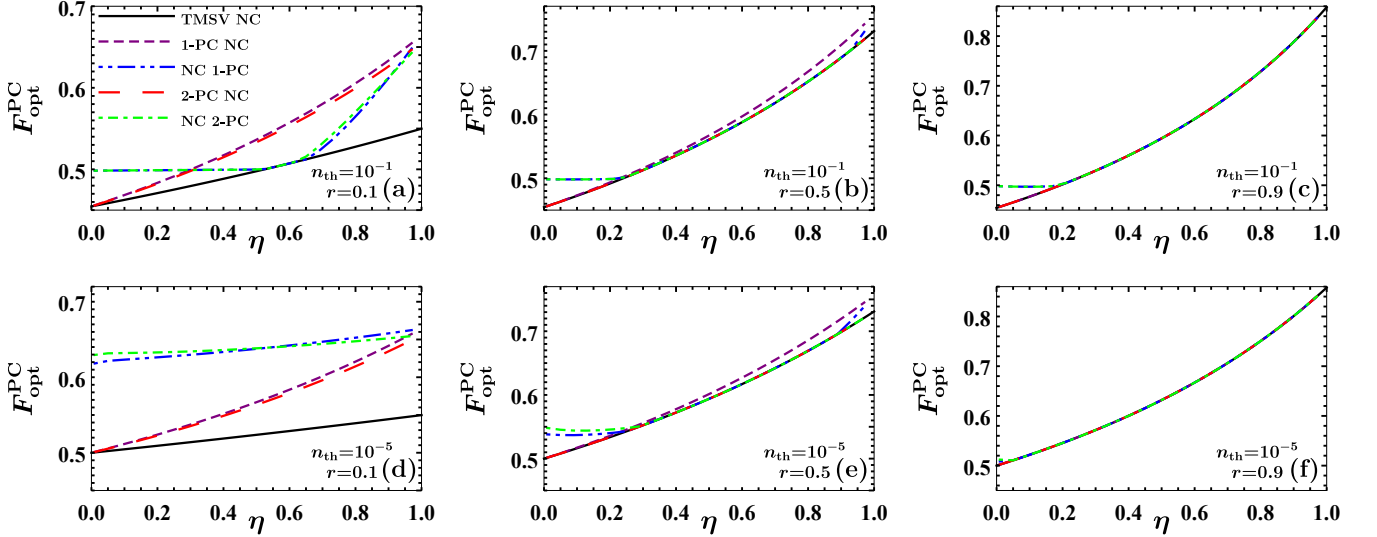}
		\caption{  Optimal fidelity as a function of transmissivity $\eta = e^{-\gamma t}$, which represents the time of interaction with the thermal bath, for different squeezing values of the TMSV state and   thermal photon number ($n_{\text{th}}$). The   fidelity has been   maximized   with respect to the transmissivity $(T)$ of the beam splitter that implements the photon catalysis operation. }
		\label{sc_grid_optF_vs_eta}
	\end{figure*}

	Among the three NG operations, we find that the PS and PC operations are advantageous for quantum teleportation as they improve the fidelity over the TMSV~NC resource state. 
	However, the PA operations
	turn out to be detrimental in this regard, and therefore, we do not explicitly show its results in this article. For the PS and PC operations, we optimize the fidelity with the transmissivity T of the beam splitter involved in the implementation of the corresponding NG operation [Fig.~\ref{ng}].
	Subsequently, we plot the optimized fidelity as a function of transmissivity $\eta$ for these operations, as shown in Figs.~\ref{ss_grid_optF_vs_eta} and~\ref{sc_grid_optF_vs_eta}, respectively. To gain better insight, we do this for different values of squeezing ($r$) and thermal photon number($n_{\text{th}}$).
	The results clearly show that performing either of these NG operations before   or after interaction with the noisy channel is significantly advantageous for a partial or complete range of $\eta$ values in comparison to the TMSV~NC resource state. 
	
	We first discuss the high thermal photon number case  ($n_{\text{th}}= 10^{-1}$) shown in the upper panel of Fig.~\ref{ss_grid_optF_vs_eta}(a), where we observe that the optimized fidelity corresponding to the NC 1-PS operation turns out to be $1/2$ for an initial range of $\eta$ values. Up to a certain value of $\eta$ ($\approx$ 0.37), it is better to implement NC 1-PS operation   rather than  1-PS~NC operation  as the former results in more fidelity. This value of $\eta$ is characterized by the crossover point among the optimal fidelity curves for these respective operations. Beyond this point, 1-PS~NC  results in better fidelity than the NC~1-PS operation.   It is worth mentioning that for the regions of $\eta$ for which 
	NC~1-PS operation is more beneficial than its counterpart, the fidelity corresponding to NC~1-PS has a constant value of $1/2$. This implies that such a resource state provides no quantum advantage as such a fidelity value can also be achieved by means of   classical  ``measure-and-prepare" strategy~\cite{Braunstein-jmo-2000,Braunstein-pra-2001}.

	We further observe that as the squeezing $r$ of the TMSV state is increased, the $\eta$ for which the crossover occurs and the range of $\eta$ values for which NC~1-PS operation has a constant value of optimized fidelity (0.5), reduce. Similar behavior is observed for the 2-PS~NC and  NC~2-PS operations.
	
	We now turn to the analysis of the  fidelity curves for PC operation in the high-temperature limit ($n_{\text{th}}=10^{-1}$), as shown in the upper panel of Fig.~\ref{sc_grid_optF_vs_eta}. As we can see, the PC operations have     more or less similar dynamics as that of the PS operation. However, we notice a change in the magnitude of $\eta$ till which a constant fidelity of 0.5   is achieved and the value  at which crossover occurs between PC~NC and NC~PC operations.

	On decreasing the thermal photon number of the bath to $n_{\text{th}}= 10^{-5}$   as shown in the lower panels of  Fig.~\ref{ss_grid_optF_vs_eta}, the plots for PS~NC and NC~PS operations become indistinguishable.   We note that the fidelity is maximized in the unit transmissivity limit. We also find that the fidelity expressions for PS~NC and NC~PS operations on TMSV resource states become the same in the limit $T \rightarrow 1$   with $n_{\text{th}}= 0$.     While the PS and PC operation cases have similar behavior for high thermal photon number, the behavior of PS and PC operations is completely different for low thermal photon number. As shown in Fig.~\ref{sc_grid_optF_vs_eta}(d), for $r=0.1$, the PC~NC and NC~PC operations have no crossover for the entire range of transmissivity ($\eta$) values and applying the PC operation after the noisy channel is beneficial for this entire range. As the value of squeezing is increased to $r=0.5$, a crossover is obtained. Prior to crossover point, the NC~PC operation is somewhat advantageous than PC~NC. After the crossover point, the 1-PC~NC operation yields a slightly higher fidelity than     NC~1-PC operation. However, the difference in fidelity for the 2-PC~NC and NC~2-PC operation is negligible. When very high squeezing values are used, as shown for $r=0.9$, the variation of fidelity amongst the different operations is practically indistinguishable.
	
	\begin{figure*}[t] 
		\centering
		\includegraphics[scale=1]{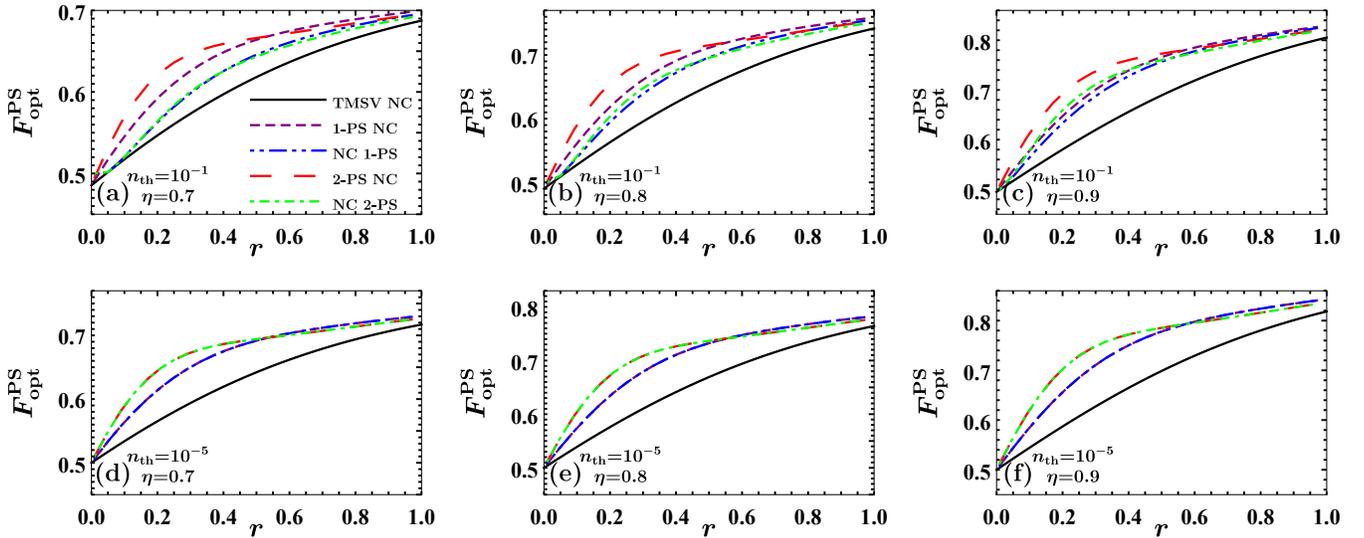}
		\caption{ Optimal fidelity as a function of squeezing parameter for different values of   thermal photon number ($n_{\text{th}}$), and  transmissivity ($\eta$) of the beam splitter that characterises the noisy channel . The   fidelity has been   maximized   with respect to the transmissivity $(T)$ of the beam splitter that implements the photon subtraction operation.  }
		\label{ss_grid_optF_vs_r}
	\end{figure*}

	We now turn up to the variation of optimized fidelity as a function of the squeezing parameter (for different values of $n_{\text{th}}$ and $\eta$) when the NG operation in question is PS   (Fig.~\ref{ss_grid_optF_vs_r}). We again observe that performing the PS   operation is advantageous for quantum teleportation as it enhances the fidelity as compared to that provided by the TMSV~NC resource state. As shown in the upper panel of Fig.~\ref{ss_grid_optF_vs_r}, the PS~NC operations yield better fidelity as compared to the NC~PS operations for almost the entire range of $r$ values when the thermal photon number is high ($n_{\text{th}}=10^{-1}$). When the thermal photon number is low ($n_{\text{th}}=10^{-5}$), the difference between the fidelity curves corresponding to PS~NC and NC~PS operations is negligible.
	\begin{figure*}[t] 
		\centering
		\includegraphics[scale=1]{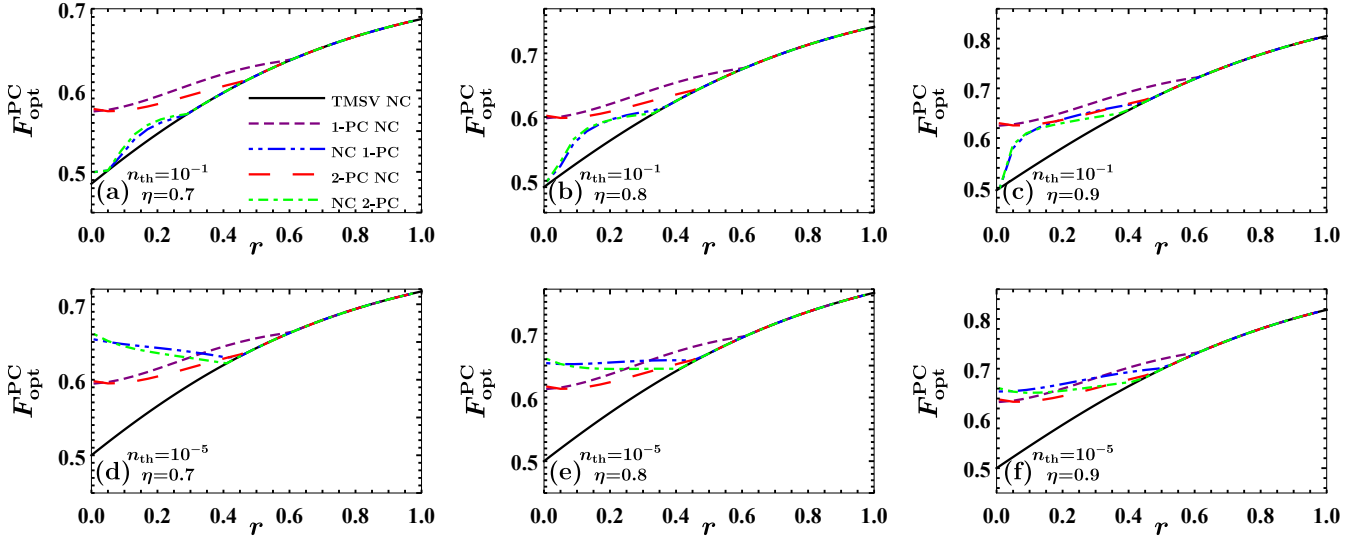}
		\caption{ Optimal fidelity as a function of squeezing parameter for different values of   thermal photon number ($n_{\text{th}}$), and  transmissivity ($\eta$) of the beam splitter that characterises the noisy channel . The   fidelity has been   maximized   with respect to the transmissivity $(T)$ of the beam splitter that implements the photon catalysis operation.  }
		\label{sc_grid_optF_vs_r}
	\end{figure*}

	Moving on to the case when the NG operation is PC, the variation of teleportation fidelity curves as a function of squeezing parameter $r$  (for different values of $n_{\text{th}}$ and $\eta$) are shown in Fig.~\ref{sc_grid_optF_vs_r}. We observe that at high temperature $n_{\text{th}}=10^{-1}$, both PC~NC and NC~PC operations provide an advantage over the original TMSV~NC  resource state till   certain  threshold values of squeezing  beyond which the optimal fidelities  equal to that obtained by the usage of  TMSV~NC resource state. The reason is that the fidelity expressions for both  PC~NC and NC~PC operations beyond these threshold  values of squeezing are optimized in the unit transmissivity limit, thereby reducing the NG state  to TMSV~NC state.
	In the squeezing range, when the fidelity curve corresponding to PC~NC operation does not coincide with that of TMSV~NC, the NC~PC operation turns out to be less advantageous than PC~NC operation for generating the resource state. 
	At low temperature $n_{\text{th}}=10^{-5}$, we observe a crossover between the fidelity of PC~NC and NC~PC operations. For small squeezing, NC~PC operation provides more fidelity as compared to PC~NC till a certain squeezing value beyond which the trend reverses for some range of $r$ and is finally followed by all the curves converging into the curve corresponding to the TMSV~NC resource state.

	\subsection{Teleportation of single-mode squeezed vacuum state }
	\begin{figure*}[t] 
		\centering
		\includegraphics[scale=1]{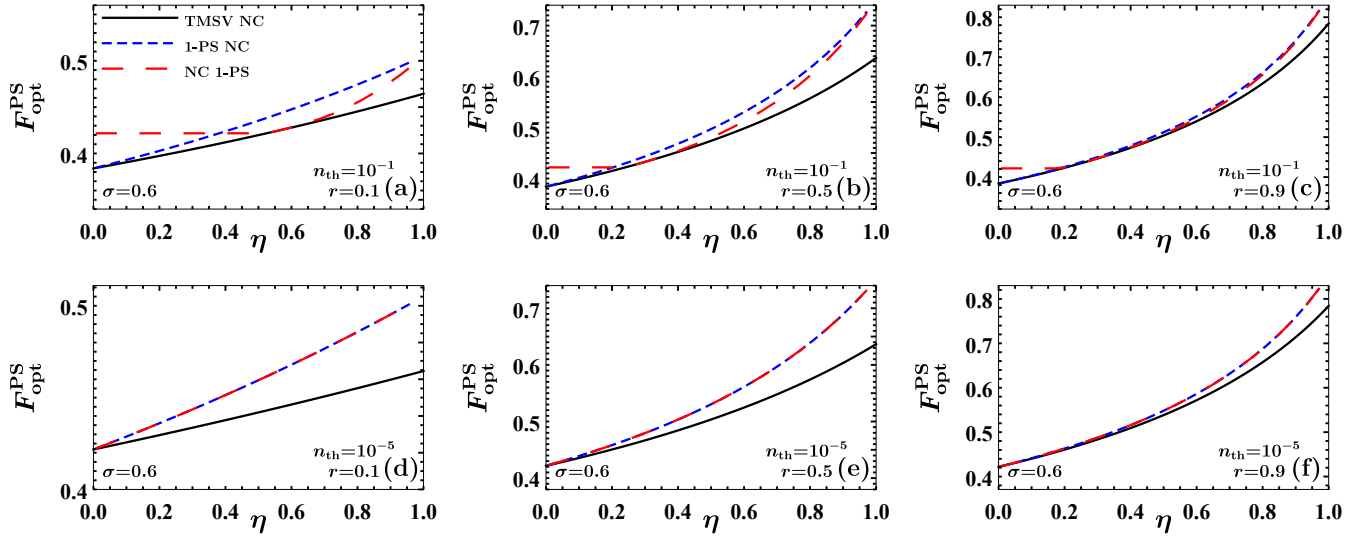}
		\caption{ Optimal fidelity as a function of squeezing parameter for different values of  thermal photon number ($n_{\text{th}}$)  and  transmissivity ($\eta$) of the beam splitter that characterizes the noisy channel . The   fidelity has been   maximized   with respect to the transmissivity $(T)$ of the beam splitter that implements the photon subtraction operation. The quantity $\sigma$ represents the squeezing of the input squeezed vacuum state to be teleported.  }
		\label{squeezed_sigma0.6_ss_grid_optF_vs_eta}
	\end{figure*}
	We now undertake the analysis of the teleportation fidelity when the input state to be teleported is a single-mode squeezed vacuum state. The squeezing   of this input state has been taken to be $\sigma=0.6$. Similar to the results of teleporting  input coherent  state, we observe that   the usage of PS or PC   operation before or after the noisy channel can provide   advantage over the TMSV~NC resource state. We first discuss the variation of the optimized fidelity with the transmissivity $\eta$ of the beam splitter that has been used to model the noisy channel. As shown in the upper panel of Fig.~{\ref{squeezed_sigma_ss_grid_optF_vs_eta}}, when the thermal photon number is high ($10^{-1}$) and the NG operation is PS, the optimized fidelity corresponding to NC~1-PS operation first remains constant up to a     
	certain threshold $\eta$ and then subsequently    starts to increase with  $\eta$. The   optimized fidelity curve corresponding to NC~1-PS operation curve   crossovers  the  1-PS~NC operation curve, with the crossover point determining the transition value of $\eta$  prior to which the NC~1-PS operation is more beneficial than 1-PS~NC operation. However, post this transition value, the 1-PS~NC operation turns out to be more advantageous in achieving higher fidelity than NC~1-PS operation. As the squeezing is increased, the values of $\eta$ for which the threshold and transition points are obtained, decrease. 
	\begin{figure*}[t] 
		\centering
		\includegraphics[scale=1]{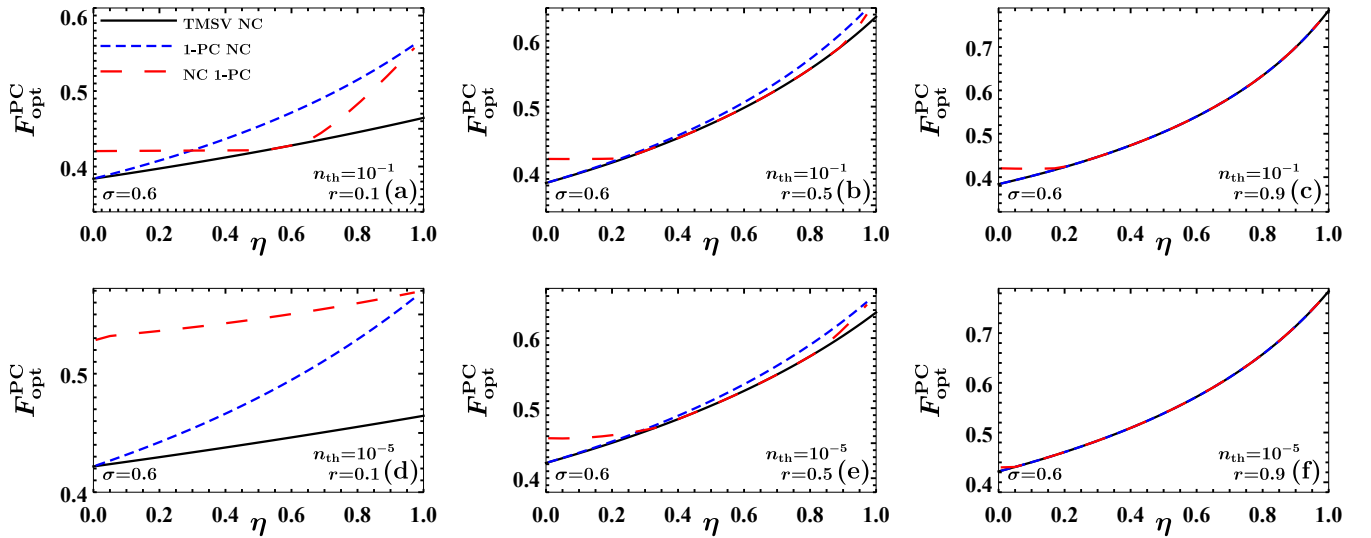}
		\caption{ Optimal fidelity as a function of squeezing parameter for different values of  thermal photon number ($n_{\text{th}}$), and  transmissivity ($\eta$) of the beam splitter that characterizes the noisy channel . The   fidelity has been   maximized   with respect to the transmissivity $(T)$ of the beam splitter that implements the photon catalysis operation. The quantity $\sigma$ represents the squeezing of the input squeezed vacuum state to be teleported.  }
		\label{squeezed_sigma0.6_sc_grid_optF_vs_eta}
	\end{figure*}

	Similar dynamics are obtained when the PS operation is replaced by the 1-PC operation with high thermal photon number ($n_{\text{th}}=10^{-1}$), as shown in the upper panel of Fig.~{\ref{squeezed_sigma0.6_sc_grid_optF_vs_eta}}. However, at high squeezing values like $r=0.9$, the threshold point and the crossover point practically coincide, beyond which there is a negligible difference amongst the fidelity curves corresponding to 1-PC~NC and NC~1-PC   and TMSV~NC.

	We now consider the low thermal photon number case as shown in the lower panel of Fig.~{\ref{squeezed_sigma0.6_ss_grid_optF_vs_eta}}. We observe that when the NG operation is PS, both 1-PS~NC and NC~1-PS operations lead to fidelity curves that are indistinguishable from each other, irrespective of the value of squeezing. Hence, both provide an almost equal advantage over the TMSV~NC resource state. The case when the associated NG operation is PC is somewhat more complicated. As depicted in the lower panel of Fig.~{\ref{squeezed_sigma0.6_sc_grid_optF_vs_eta}}, when $r=0.1$, NC~1-PC turns out to be notably advantageous over 1-PC~NC over almost the entire range of transmissivity $\eta$. At $r=0.5$, the NC~1-PC curve initially exhibits somewhat higher fidelity than 1-PC~NC but the trend soon reverses with 1-PC~NC, resulting in a slight advantage over NC~1-PC. With the squeezing being further increased to $r=0.9$, the difference between the fidelity curves is practically negligible.
	
	We again note that photon addition does not improve the teleportation fidelity over the original state in teleporting input squeezed vacuum state; therefore, we do not show the results.

\begin{figure*}[t] 
		\centering
		\includegraphics[scale=1]{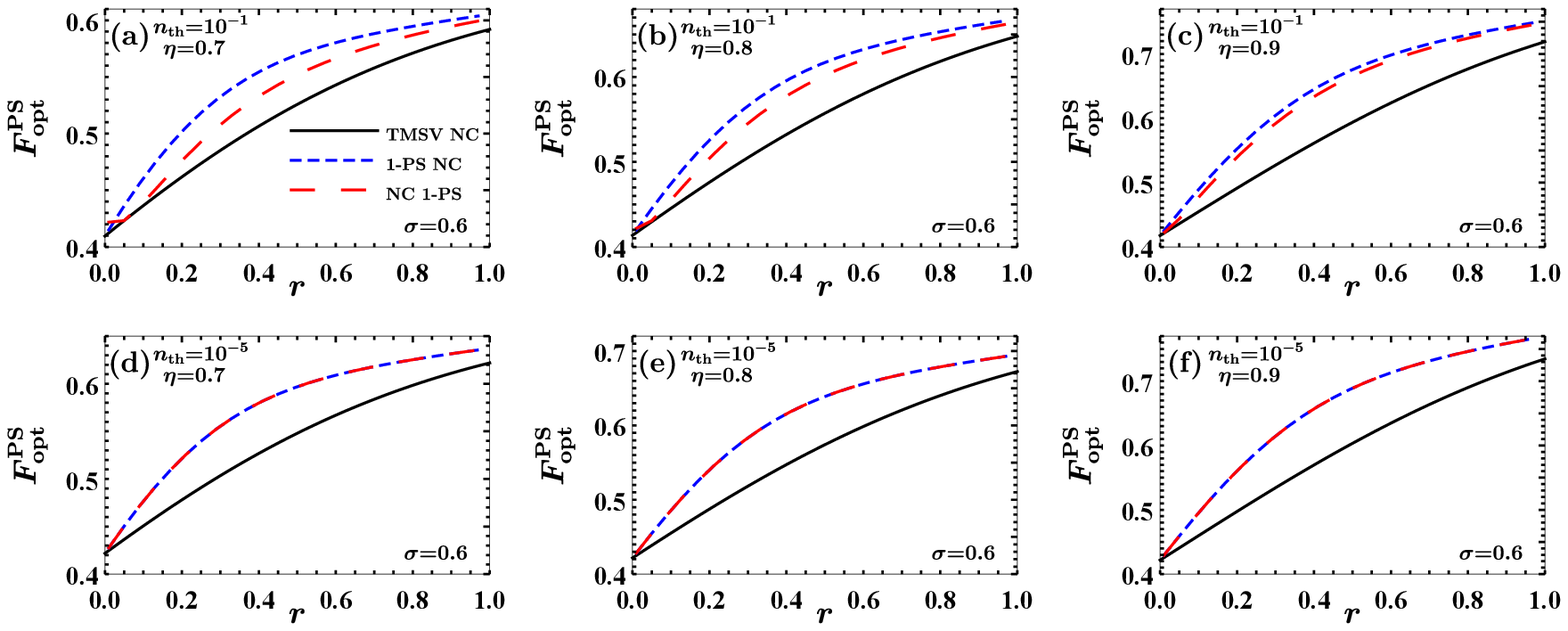}
		\caption{ Optimal fidelity as a function of squeezing parameter for different values of  thermal photon number ($n_{\text{th}}$), and  transmissivity ($\eta$) of the beam splitter that characterises the noisy channel. The   fidelity has been   maximized   with respect to the transmissivity $(T)$ of the beam splitter that implements the photon addition operation. The quantity $\sigma$ represents the squeezing of the input squeezed vacuum state to be teleported.  }
		\label{squeezed_sigma0.6_ss_grid_optF_vs_r}
	\end{figure*}

	We now move on to the study of optimized fidelity as a function of the squeezing parameter (for different values of $n_{\text{th}}$ and $\eta$)   in Fig.~\ref{squeezed_sigma0.6_ss_grid_optF_vs_r} for the case of photon subtraction.
	The relative dynamics of the   1-PS~NC, NC~1-PS and TMSV~NC are similar to that obtained when the input state was taken to be a coherent state [Fig.~\ref{ss_grid_optF_vs_r}]. It turns out that implementing   PS operation before or after noisy channel provides a significant advantage over the TMSV~NC resource state. As shown in the upper panel of Fig.~{\ref{squeezed_sigma0.6_ss_grid_optF_vs_r}}, the 1-PS~NC operation turns out to be more beneficial than NC~1-PS at high temperature ($n_{\text{th}}=10^{-1}$). In contrast, at low temperature ($n_{\text{th}}=10^{-5}$), a negligible difference exists among the optimized fidelity curves of the two operations (lower panel of Fig.~{\ref{squeezed_sigma0.6_ss_grid_optF_vs_r}}). 
	
	\begin{figure*}[t]  
		\centering
		\includegraphics[scale=1]{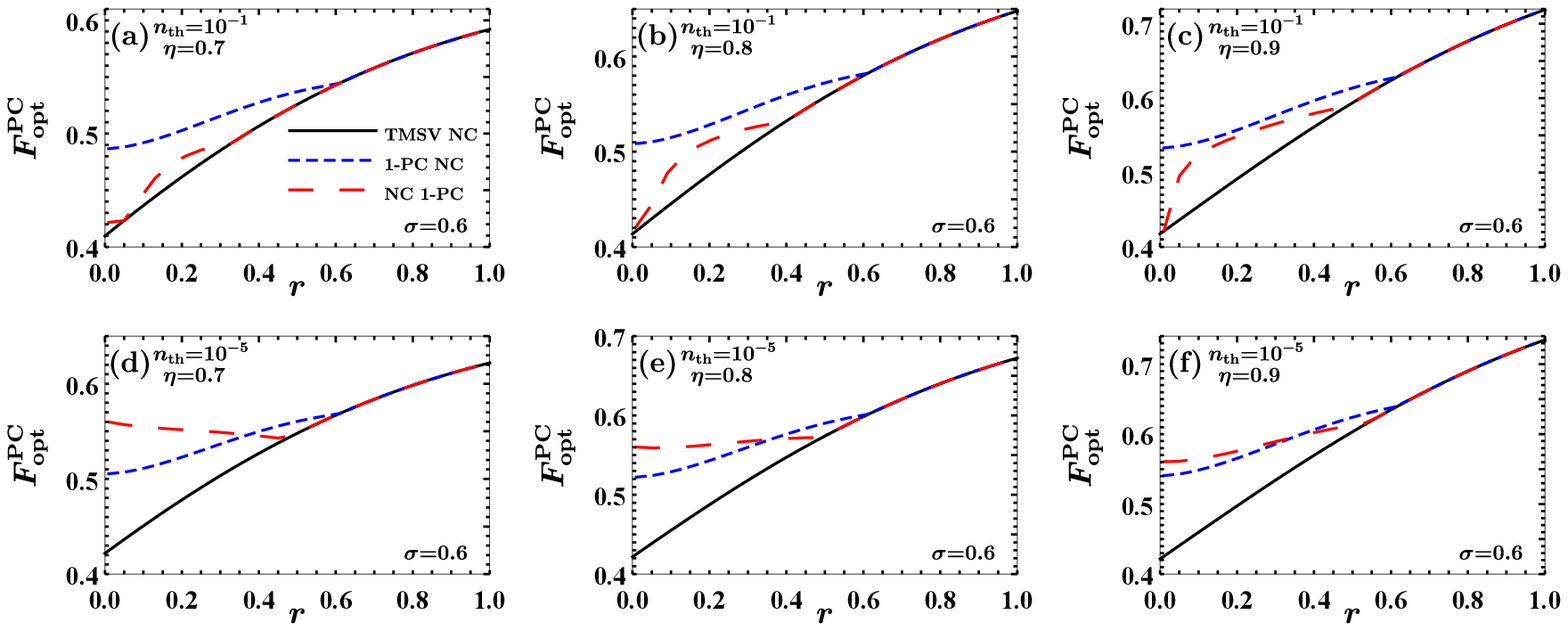}
		\caption{ Optimal fidelity as a function of squeezing parameter for different values of   thermal photon number ($n_{\text{th}}$), and  transmissivity ($\eta$) of the beam splitter that characterises the noisy channel. The   fidelity has been   maximized   with respect to the transmissivity $(T)$ of the beam splitter that implements the photon subtraction operation. The quantity $\sigma$ represents the squeezing of the input squeezed vacuum state to be teleported. }
		\label{squeezed_sigma0.6_sc_grid_optF_vs_r}
	\end{figure*}
	
	We now move on to the scenario when the NG operation is PC. The upper panel of Fig.~{\ref{squeezed_sigma0.6_sc_grid_optF_vs_r}} shows the case when the thermal photon number is high ($n_{\text{th}}=10^{-1}$). We see that for an initial range of squeezing, the 1-PC~NC operation in comparison to NC~1-PC operation, leads to the generation of a much favorable resource state. There exists a threshold squeezing  for each of the 1-PC~NC and  NC~1-PC curves beyond which,   the optimized fidelity   maximize in the unit transmissivity limit ($T\rightarrow 1$), and therefore, the non-Gaussian  state reduces to the TMSV~NC   state. At low temperature ($n_{\text{th}}=10^{-5}$), higher fidelity is obtained by the NC~1-PC  operation as compared to 1-PC~NC operation until both the curves   crossover. After the crossover point, for a small interval of $r$,  1-PC~NC results in a slightly higher fidelity than the  NC~1-PC, with both curves  finally merging with the fidelity curve corresponding to the TMSV~NC state, the reason being the same as mentioned for the high thermal photon number case.

	\section{Conclusion}
	\label{sec:conc}

	In this article, we compared two distinct scenarios in the context of CV quantum teleportation where we perform different NG operations, namely, photon subtraction, addition, and catalysis, before and after noisy channel. In the first scenario, we considered the case where the resource state is generated by performing NG operations on the TMSV state followed by interaction with noisy channel. In the second case, the NG operations are performed after the TMSV state has interacted with noisy channel.  
	The findings indicate that either of the two approaches could be more effective depending on the type of non-Gaussian operation, the initial squeezing of the TMSV state, and the thermal bath parameters.

	The current strategies of performing different NG operations before or after interaction with noisy environment enable us to lessen the detrimental effects of the environment. Such strategies have already been employed in CV QKD protocols~\cite{pra-catalysis-2021} and can be incorporated in other CV QIP protocols such as quantum metrology~\cite{crs-ngtmsv-met} and quantum illumination~\cite{Saikat}.
	We can also consider more general environmental models such as global environment~\cite{xiang-pra-2008} and non-Markovian thermal environment~\cite{pablo-prl-2008,Non-Markovian} and understand their implications on different CV QIP protocols.

	\section*{Acknowledgement}
	This is the fifth article in a publication series
	written in the celebration of the completion of 15 years of IISER Mohali.	  C.K. acknowledges the financial support from 
	{\bf DST/ICPS/QuST/Theme-1/2019/General} Project number {\sf Q-68}.

	%%%%%%%%%%%%%%%%%%%%%%%%%%%%%%%%%%%%%%%%%%%%%

	%%%%%%%%%%%%%%%%%%%%%%%%%%%%%%%%%%%%%%%%%%%%%
	
 %\bibliography{references}
	

	%%%%%%%%%%%%%%%%%%%%%%%%%%%%%%%%%%%%%%%%%%%%%%%%%	
	
\end{document}